\documentclass[%
reprint,
superscriptaddress,
showpacs,
amsmath,amssymb,
aps,
pra,
]{revtex4-1}

\usepackage{amsmath}
\usepackage{changes}
\usepackage{graphicx}
\usepackage{braket}
\usepackage{amssymb}
\usepackage{float}
\usepackage{color}
\usepackage{comment}
\usepackage[colorlinks=true,linkcolor=blue,anchorcolor=blue,citecolor=blue,urlcolor=blue]{hyperref}
\usepackage{bm}
\begin{document}
\preprint{APS/123-QED}

\title{Benchmarking Atomic Ionization Driven by Strong Quantum Light}


\author{Yi-Jia Mao}
\thanks{These authors contributed equally to this work.}
\affiliation{State Key Laboratory of Dark Matter Physics, Key Laboratory for Laser Plasmas (Ministry of Education) and School of Physics and Astronomy, Collaborative Innovation Center for IFSA (CICIFSA), Shanghai Jiao Tong University, Shanghai 200240, China}
\affiliation{Tsung-Dao Lee Institute, Shanghai Jiao Tong University, Shanghai 201210, China}
\author{En-Rui Zhou}
\thanks{These authors contributed equally to this work.}
\affiliation{State Key Laboratory of Dark Matter Physics, Key Laboratory for Laser Plasmas (Ministry of Education) and School of Physics and Astronomy, Collaborative Innovation Center for IFSA (CICIFSA), Shanghai Jiao Tong University, Shanghai 200240, China}
\author{Yang Li}
\email{liyang22@sjtu.edu.cn}
\affiliation{State Key Laboratory of Dark Matter Physics, Key Laboratory for Laser Plasmas (Ministry of Education) and School of Physics and Astronomy, Collaborative Innovation Center for IFSA (CICIFSA), Shanghai Jiao Tong University, Shanghai 200240, China}
\author{Pei-Lun He}
\email{peilunhe@sjtu.edu.cn}
\affiliation{State Key Laboratory of Dark Matter Physics, Key Laboratory for Laser Plasmas (Ministry of Education) and School of Physics and Astronomy, Collaborative Innovation Center for IFSA (CICIFSA), Shanghai Jiao Tong University, Shanghai 200240, China}
\author{Feng He}
\email{fhe@sjtu.edu.cn}
\affiliation{State Key Laboratory of Dark Matter Physics, Key Laboratory for Laser Plasmas (Ministry of Education) and School of Physics and Astronomy, Collaborative Innovation Center for IFSA (CICIFSA), Shanghai Jiao Tong University, Shanghai 200240, China}
\affiliation{Tsung-Dao Lee Institute, Shanghai Jiao Tong University, Shanghai 201210, China}

\date{\today}

\begin{abstract}
The recently available high-intensity quantum light pulses provide novel tools for controlling light-matter interactions. However, the rigor of the theoretical frameworks currently used to describe the interaction of strong quantum light with atoms and molecules remains unverified. Here, we establish a rigorous benchmark by solving the fully quantized time-dependent Schr\"{o}dinger equation for an atom exposed to bright squeezed vacuum light. Our \textit{ab initio} simulations reveal a critical limitation of the widely used $Q$‑representation: although it accurately reproduces the total photoelectron spectrum after tracing over photon states, it completely fails to capture the electron-photon joint energy spectrum. To overcome this limitation, we develop a general theoretical framework based on the Feynman path integral that properly incorporates the electron-photon quantum entanglement. Our results provide both quantitative benchmarks and fundamental theoretical insights for the emerging field of strong-field quantum optics.

\end{abstract}
\maketitle

In conventional strong-field physics, laser pulses are predominantly described as Glauber coherent states~\cite{qrep_hhg1}, an approximation that neglects the quantum fluctuations of light. Semiclassical theories based on the time-dependent Schr\"{o}dinger equation (TDSE) that treat the laser pulses as external classical fields~\cite{hhg3,sfa,review1,review2,qprop,qpc} have successfully underpinned landmark discoveries in attosecond science, including high-harmonic generation (HHG)~\cite{hhg1,hhg2,hhg3,hhg4}, above-threshold ionization (ATI)~\cite{ati1,ati2,ati3}, and nonsequential double ionization (NSDI)~\cite{nsdi1,nsdi2,nsdi3}.
In parallel, quantum optics has long exploited nonclassical states of light, such as Fock states, squeezed states, and thermal states, for applications in quantum metrology, quantum communication, and foundational tests of quantum theory~\cite{quantum_optics,quaninf1,quanmetro,quanop1}. Recent experimental breakthroughs have enabled the generation of ultrashort quantum light pulses at intensities reaching $10^{12}$-$10^{13}\,\mathrm{W/cm^2}$, comparable to the Coulomb field that binds valence electrons in atoms~\cite{hhg_bsv_solid,exp_bsv_1,exp_bsv_2,exp_bsv_3}. 
This development marks a convergence between quantum optics and strong-field physics, giving rise to the emerging discipline of strong-field quantum optics. Accordingly, a compelling question arises: how do the photon statistics and quantum fluctuations of nonclassical fields influence strong-field electronic dynamics?

At present, theoretical treatments of strong-field quantum-light-matter interactions rely heavily on the generalized $P$-representation~\cite{drummond1980generalised}, from which the $Q$-representation is obtained by taking an appropriate limit for practical computations~\cite{qrep_hhg1,qrep_hhg2,qati1,qdouble,qati2}. Within this framework, one can derive quantities such as HHG~\cite{hhg_bsv_solid,qrep_hhg1,qrep_hhg2}, ATI spectra~\cite{qati1,qati2,qati3} and kinetic energy release in dissociation~\cite{dis}. While the validity of the $Q$-representation has been verified for HHG using the von Neumann lattice method~\cite{full_quantum_hhg}, the $Q$‑representation faces conceptual challenges regarding which operator corresponds to the experimental observables, leading to inconsistent results in the literature~\cite{hhg_bsv_solid,qrep_hhg1,qrep_hhg2,qati1,qati2,qati3,qhhg1,qhhg2}. Crucially, the validity of the $Q$‑representation for general dynamical systems remains unclear and calls for systematic benchmarking.

In this Letter, we present an \textit{ab initio} numerical simulation of the TDSE for the complete atom-photon system in the strong‑field regime, where both the electronic and optical degrees of freedom are treated quantum mechanically. This simulation establishes a rigorous benchmark for strong-field quantum optics. We find that predictions from the widely used $Q$-representation are consistent for the total electron signals after tracing over the photon degrees of freedom. Crucially, however, the photon statistics of the driving field deviate significantly from the predictions of the $Q$-representation. This discrepancy motivates a fundamental reformulation of the problem. To this end, we rigorously derive a general expression starting from the Feynman path integral for the coupled atom-photon system. Our formulation not only yields a more accurate theoretical tool but also provides clear physical insight into the origin of the discrepancy. By focusing on the photon-state correlated photoelectron energy spectrum, we establish a general picture of the limits and validity of the $Q$-representation in describing the interaction between intense quantum light and atoms.

\begin{figure*}
\centering
\includegraphics[width=1\textwidth]{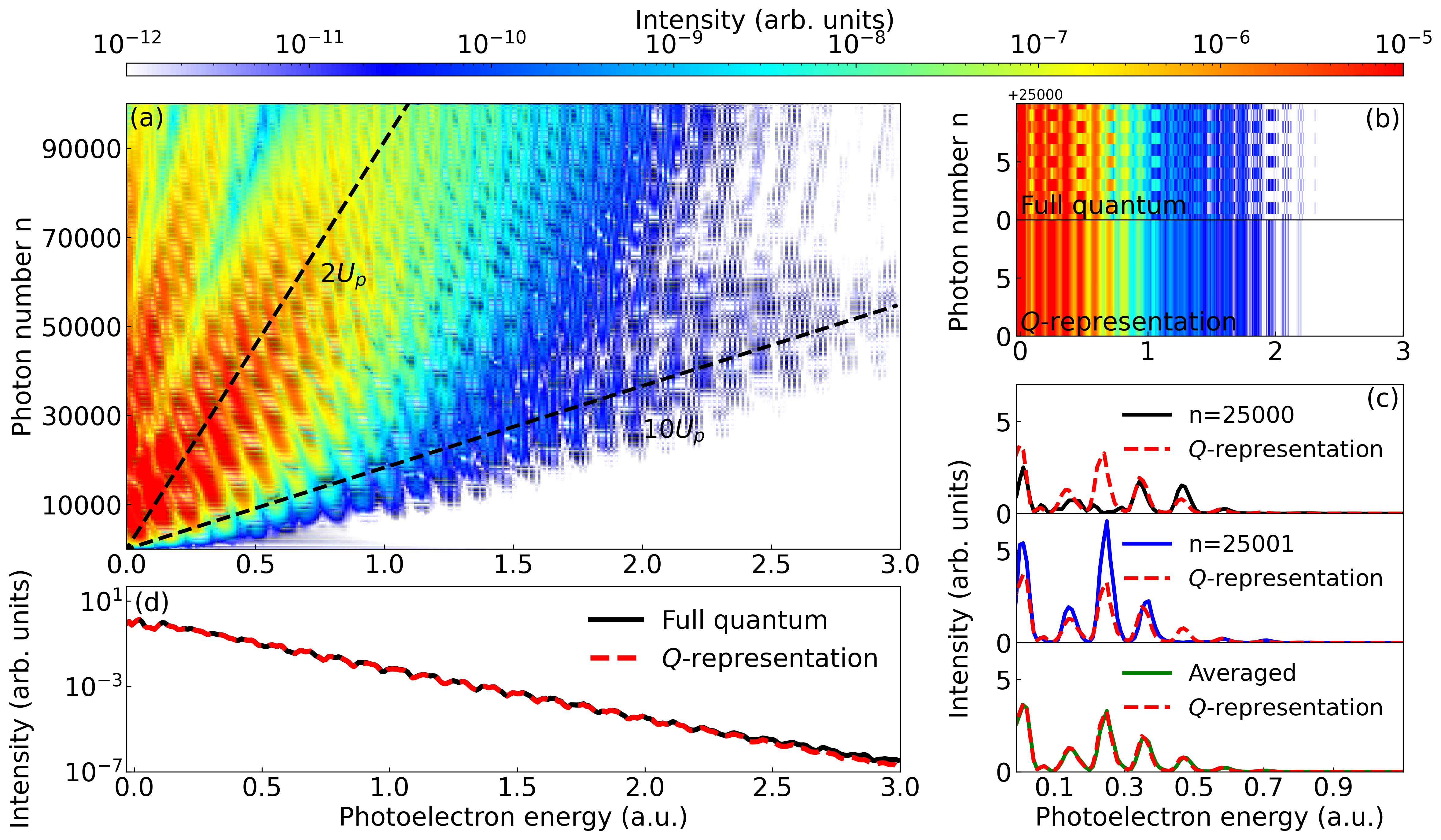}
\caption{(a) Joint photon-photoelectron energy spectrum obtained from the full quantum TDSE simulation. (b) Magnified view of the joint energy spectrum for photon numbers in the range $25000-25010$. The upper panel shows the results from the full quantum TDSE simulation, while the lower panel shows the corresponding results from the $Q$-representation method. (c) Photoelectron energy spectra correlated with specific Fock states. The upper and middle panels show the spectra for $n=25000$ and $n=25001$, respectively, while the lower panel shows their averaged result. Solid and dashed curves denote the results from the full quantum TDSE and the $Q$-representation methods, respectively. (d) Total photoelectron energy spectra calculated using the full quantum TDSE simulation (black solid curve) and the $Q$-representation method (red dashed curve).}
\label{Fig1}
\end{figure*}
We begin by numerically solving the TDSE for a one-dimensional model hydrogen atom interacting with a quantized photon field. The total wave function of the coupled atom-photon system satisfies (atomic units are used throughout unless stated otherwise)
 \begin{equation}
 i\dfrac{\partial}{\partial t}\ket{\Psi(t)}=\hat{H}\ket{\Psi(t)}.
 \label{fullqseq}
 \end{equation}
We expand $\ket{\Psi(t)}$ in the direct product basis
\begin{equation}
\ket{\Psi(t)}=\sum_{n=0}^{N_\gamma}\int \mathrm{d}x~ c_{x,n}(t)\ket{x}\otimes\ket{n},\label{fullpsi}
\end{equation}
where $\ket{x}$ and $\ket{n}$ denote the electron position ket and the photon Fock state, respectively. The Hamiltonian in the interaction picture is given by $\hat{H}=\hat{H}_A+\hat{H}_{\rm int}$, where
$\hat{H}_A=-\dfrac{1}{2}\dfrac{\mathrm{d}^2}{\mathrm{d}x^2}-\dfrac{1}{\sqrt{x^2+2}}$
is the field-free atomic Hamiltonian, and
$\hat{H}_{\rm int}=i f(t) \epsilon_V \hat{x}(\hat{a}e^{-i\omega t}-\hat{a}^\dagger e^{i\omega t})$ describes the atom-photon interaction in the dipole approximation. Here, $\hat{a}$ and $\hat{a}^\dagger$ are the photon annihilation and creation operators, $f(t)$ is the pulse envelope, $\omega$ the angular frequency, and $\epsilon_V=\sqrt{{2\pi\omega}/{V}}$ the single-photon field amplitude associated with the pulse volume $V$~\cite{qV}. 
The coupled wave function is propagated using the Crank-Nicolson method~\cite{cn}, and the photoelectron energy spectra (PES) are extracted at the end of the pulse using a window-operator technique~\cite{window1,window2,window3}. Additional computational details are provided in Sec.~I and II in the \textit{Supplemental Materials}~\cite{supp}.

Having established the numerical framework for solving the TDSE with a fully quantized photon field, we now assess the performance of the $Q$-representation approach~\cite{qrep_hhg1,qrep_hhg2,qati1,qdouble,qati2} to compare it with the full quantum TDSE simulation. The total density matrix given by the $Q$-representation is
\begin{equation}
    \rho^{(Q)}(t)=\int\mathrm{d}^2\alpha\,Q(\alpha)\,|\psi_e(t;\alpha)\rangle\langle\psi_e(t;\alpha)|\otimes|\alpha\rangle\langle\alpha|,
    \label{rhoQ}
\end{equation}
where $|\alpha\rangle$ is a coherent state, $Q(\alpha)={\langle\alpha|\phi_{\gamma}\rangle\langle\phi_{\gamma}|\alpha\rangle}/{\pi}$ the $Q$-representation of the initial photon state $|\phi_{\gamma}\rangle$, and $|\psi_e(t;\alpha)\rangle$ the electron state driven by the classical laser pulse with coherent-state amplitude $\alpha$,
\begin{equation}
    \begin{aligned}
 i\dfrac{\partial}{\partial t}|\psi_e(t;\alpha)\rangle=\left[\hat{H}_A+2\hat{x}|\alpha|\epsilon_V\sin(\omega t-\theta_\alpha)f(t)\right]|\psi_e(t;\alpha)\rangle
    \end{aligned},
\end{equation}
where $\theta_{\alpha}=\arg[\alpha]$ is the carrier-envelope phase. 

As a representative scenario, we consider strong-field ionization driven by a bright squeezed vacuum (BSV) at a wavelength of 400~nm. The BSV field has a peak intensity of $10^{14}~\mathrm{W/cm^2}$. The pulse envelope is trapezoidal, consisting of one optical cycle of ramp‑up, two cycles at full intensity, and one cycle of ramp‑down. For a given laser intensity in the quantum-field model, the effective electric-field amplitude scales with the single-photon amplitude as $E_0 = 2\epsilon_V \sinh r$, while the average photon number is $\bar{n} = \sinh^2 r$. Unless stated otherwise, we take the squeezing parameter $r=5.3$ to ensure convergence within a tractable Fock basis. Convergence is verified in Sec.~II in the \textit{Supplemental Materials}~\cite{supp}.

To visualize the complete quantum dynamics, we first examine the joint electron-photon energy spectrum by calculating the PES associated with a specific correlated photon-number state. The resulting spectra are shown in Fig.~\ref{Fig1}(a). Several distinct patterns emerge, highlighting that both the electron and light field are modulated during the interaction. The ATI peaks appear as stripes spaced by the photon energy $\omega$. Notably, these stripes tilt from top‑left to bottom‑right, consistent with the ponderomotive energy shift $U_p^{(n)} = {\epsilon_V^2}(2n+1)/{2\omega^2}$ of the final state $|E\rangle\otimes|n\rangle$~\cite{supp}. Two well-defined energy cutoffs emerge at $E=2U_p^{(n)}$ and $E=10U_p^{(n)}$, corresponding to direct ionization and rescattering ionization, respectively,  in agreement with well-established semiclassical theories~\cite{milovsevic2006above}. 
Since the ponderomotive shift  $U_p^{(n)}$ associated with each Fock state increases linearly with $n$, these cutoffs appear as two straight lines as marked in  Fig.~\ref{Fig1}(a). Moreover, in the region $E<2U_p^{(n)}$, two pronounced local minima emerge as additional linear features. These minima arise from intra-cycle interference~\cite{intra}, which also scales proportionally with $U_p^{(n)}$, and thus with the photon number $n$.

Beyond the classical strong-field features discussed above, the fully quantized simulation uncovers additional structures. In particular, Fig.~\ref{Fig1}(a) exhibits dense horizontal fringes, which are more clearly presented in a magnified view in the upper panel of Fig.~\ref{Fig1}(b). The results show subtle photon-number-dependent modulations of the ionization channels. To assess whether such quantum-correlated features can be reproduced by semiclassical quantum light modeling, we calculate the corresponding joint spectrum using the commonly adopted $Q$-representation method~\cite{supp}, as shown in the lower panel of Fig.~\ref{Fig1}(b). A clear discrepancy emerges: The full quantum results reveal that photoelectron signals correlated with even and odd Fock states exhibit distinct behaviors. In particular, different ATI peaks are predominantly associated with either even or odd photon-number states. In stark contrast, although calculations based on the $Q$-representation capture the overall envelope, the spectrum appears continuous and lacks the even-odd modulation associated with the Fock states. This comparison immediately demonstrates that while the $Q$-representation may approximate averaged observables, it fails to capture the granular quantum correlations and photon statistics inherent in strong-field quantum optics. 
Figure~\ref{Fig1}(c) highlights this point more explicitly through a direct comparison between the two approaches for representative even ($n=25000$) and odd ($n=25001$) photon-number components. The $Q$-representation overestimates the even-$n$ contributions while underestimating the odd-$n$ ones. Remarkably, however, once these components are averaged, the two methods become nearly indistinguishable. 

Motivated by this observation, we now compare the total PES obtained by summing over all final photon-number states. As shown in Fig.~\ref{Fig1}(d), the full quantum TDSE and $Q$-representation results become essentially identical. This agreement arises because the total PES integrates over the photon degree of freedom, thereby eliminating photon-number-resolved information and masking the deficiencies of the $Q$-representation in reproducing the electron-photon entangled dynamics. This is a critical finding for strong-field quantum optics, as it validates the use of the computationally simpler $Q$-representation for systems where only the electron's final kinetic energy distribution is measured.
We emphasize, however, that unlike the joint photon-photoelectron energy spectrum, which exhibits clear ATI peaks and cutoff structures, the total PES displays strongly blurred ATI modulations and barely visible cutoffs. This loss of structure has recently been reported experimentally in electron energy spectra generated at a metal nanotip using quantum light~\cite{needle}. Here, we show that such broadening originates from quantum photon-number fluctuations intrinsic to the driving field, which, in principle, can be experimentally measured in strong-quantum-field ionization of atoms.

\begin{figure}
\centering
\includegraphics[width=0.48\textwidth]{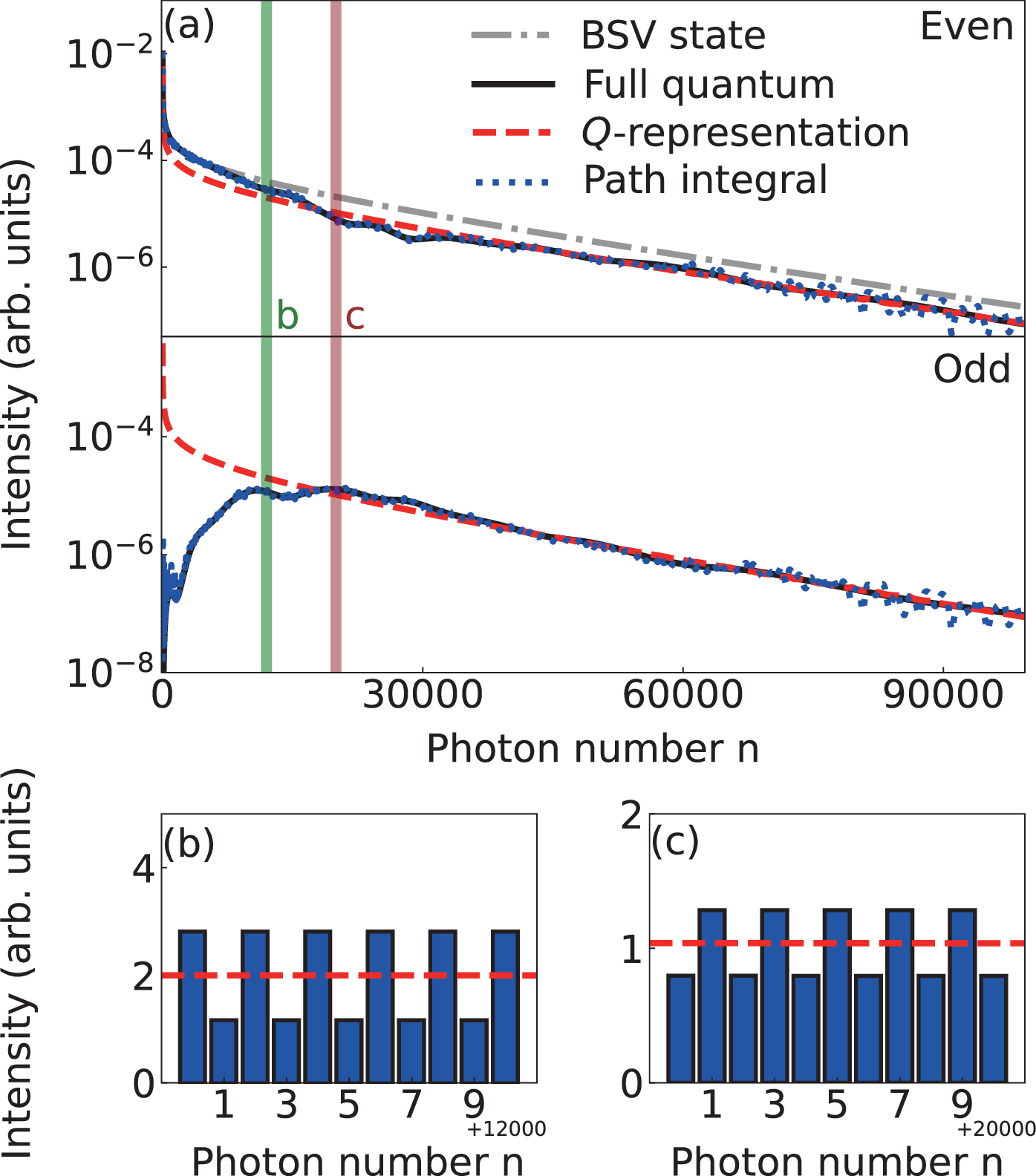}
\caption{(a) Comparison of the photon-number distributions of $400$~nm BSV light after interaction with a 1D hydrogen atom, calculated using the full quantum TDSE (black solid line), the $Q$-representation (red dashed line), and the path integral method (blue dotted line). The photon-number distribution of the incident BSV light is shown by the gray dash-dotted line. 
The upper and lower panels show the distributions for even and odd photon-number states, respectively. (b) Magnified photon-number distribution from the full quantum TDSE simulation in the range $12000-12010$. The red dashed line shows the result obtained using the $Q$-representation. (c) Same as (b), but in the range $20000-20010$.}
\label{Fig2}
\end{figure}


To pinpoint the origin of the breakdown of the $Q$-representation method in the joint spectra, we examine the underlying photon-number statistics.
In the full quantum TDSE simulation, the photon-number distribution after the light-atom interaction is computed as $ P_\gamma^{(F)}(n, t)\equiv \int \mathrm{d}x \left | \left< x, n \right|  \left. \Psi(t)\right> \right |^2$, as shown by the black solid curves in Fig.~\ref{Fig2}(a). The upper and lower panels show the distributions for even and odd photon-number states, respectively. Initially, the BSV state (gray dash-dotted curve) exhibits strict even photon-number parity. After interacting with the atom, however, the light state becomes a mixed state containing both even and odd photon contributions. The zoom-in views around $n\approx 12000$ and $20000$, shown in Figs.~\ref{Fig2}(b) and~\ref{Fig2}(c), further reveal that the local dominance of even versus odd photon numbers varies across the distribution. This modulation directly reflects the entangled nature of electrons and photons during the interaction, which is inherently lost in semiclassical approximations.

As a comparison, we also calculate the photon statistics in the $Q$-representation. The photon statistics predicted by the $Q$-representation are obtained as $P^{(Q)}_{\gamma}(n,t)= \int \mathrm{d}^2\alpha\, Q(\alpha)\,|\langle n|\alpha\rangle|^2$ by tracing out the degree of the electron from the density matrix shown by Eq.~\eqref{rhoQ}. Crucially, within this framework, the photon-number distribution remains time-independent throughout the entire light-matter interaction, as shown by the red dashed curves in Fig.~\ref{Fig2}(a). No parity mixing or dynamical redistribution occurs. As a result, the $Q$-representation does not capture the light-field modulation observed in the full quantum TDSE simulation. It only reproduces the photon-averaged outcome, consistent with its behavior in the joint energy spectrum discussed above.

To understand how the electron-photon entanglement leads to the breakdown, we use the Feynman path integral approach \cite{kleinert2006path,cruz2023forward} and derive an approximate wave function that retains the correlated dynamics presented in the full quantum theory. The central results are summarized below, with additional derivation details provided in Sec.~IV in the \textit{Supplemental Materials}~\cite{supp}.
By expressing the state vector evolution via the Feynman path integral with coherent states as the photon-field basis, we analytically evaluate the coherent-state integration, yielding
\begin{equation}
    \begin{aligned}
        |\Psi(t_f)\rangle=&\int\frac{\mathrm{d}^2\alpha_f}{\pi}\int\frac{\mathrm{d}^2\alpha_i}{\pi} \int \mathrm{d}x_f\int \mathrm{d}x_i\int^{x_f}_{x_i}\mathcal{D}x\mathcal{D}p \,\\
        &e^{ i\int_{t_i}^{t_f}\mathrm{d}t\,\left[p\dot{x}-H_A-if(t)\epsilon_Vx\left(\alpha(t)e^{-i\omega t}-\alpha_f^{*}e^{i\omega t}\right)\right]}\\    \times&\langle\alpha_f|\alpha_i\rangle\langle\alpha_i|\phi_{\gamma}\rangle\langle x_i|\psi_e(t_i)\rangle ~|x_f\rangle\otimes |\alpha_f\rangle
        \label{pathint},
    \end{aligned}
\end{equation}
where $\alpha(t)=\alpha_i-\epsilon_V\int_{t_i}^t \mathrm{d}t'f(t')x(t')e^{i\omega t'}$ is the solution of the Maxwell equation $\dot{\alpha}(t)=-\epsilon_Vx(t)f(t)e^{i\omega t}$, which describes how the reaction of electron radiation changes the field of light. In the limit $\epsilon_V \to 0$, the backaction on the light field described by $\alpha(t)$ becomes negligible, allowing one to set $\alpha(t) \to \alpha_i$; consequently, $\alpha_f \to \alpha_i$.Thereby, we obtain an approximate solution to the full quantum TDSE shown in Eq.~\eqref{fullqseq} as
\begin{equation}
    \begin{aligned}
    |\Psi(t)\rangle=\int\frac{\mathrm{d}^2\alpha}{\pi}\langle\alpha|\phi_{\gamma}\rangle|\psi_e(t;\alpha)\rangle\otimes|\alpha\rangle
        \label{psi},
    \end{aligned}
\end{equation}
which can be applied to calculate the physical observables. Equation \eqref{psi} has a clear physical interpretation: The initial photon state $|\phi_{\gamma}\rangle$ is expanded in the coherent-state basis $|\alpha\rangle$, where each component independently steers the electron. This formulation inherently encodes the electron-photon entanglement, and the resulting system's density matrix takes the form
\begin{equation}
\begin{aligned}
\rho^{(R)}(t)
=&
\int \frac{\mathrm{d}^2\alpha}{\pi}\,
     \frac{\mathrm{d}^2\beta}{\pi}\,
     R(\alpha^{*},\beta)\,
     e^{-\frac{1}{2}\left(|\alpha|^2 + |\beta|^2\right)}
\\
&\times
\left(|\psi_e(t;\alpha)\rangle\langle\psi_e(t;\beta)|\right) \otimes\left(|\alpha\rangle\langle\beta|\right),
\label{rho}
\end{aligned}
\end{equation}
where
$R(\alpha^{*},\beta)=\langle\alpha|\phi_{\gamma}\rangle\langle\phi_{\gamma}|\beta\rangle\, e^{\frac{1}{2}(|\alpha|^2+|\beta|^2)}$
is the $R$‑representation. Partial tracing out the electron yields the photon density matrix composing the time-dependent factor $\langle\psi_e(t;\beta)|\psi_e(t;\alpha)\rangle$. The corresponding photon-number distribution takes the form
\begin{equation}
    \begin{aligned}
        P^{(R)}_{\gamma}(n,t)=\int \mathrm{d}x\,\left|\int\frac{\mathrm{d}^2\alpha}{\pi}\langle n|\alpha\rangle\langle\alpha|\phi_{\gamma}\rangle \langle x|\psi_e(t;\alpha)\rangle\right|^2.
    \end{aligned}\label{PR}
\end{equation} 
Neglecting the off-diagonal terms in $R(\alpha^{*},\beta)$ reduces it to the $Q$-representation given in Eq.~\eqref{rhoQ}~\cite{supp}, thereby eliminating the time evolution of the photon-number distribution. The blue dashed curves in Fig.~\ref{Fig2}(a) show the photon-number distributions calculated by Eq.~\eqref{PR}. It agrees very well with the full quantum TDSE results, assuring that the approximate state in Eq.~\eqref{psi} accurately describes both the electron and photon subsystems. It is instructive to examine why this approximation still succeeds even though Eq.~\eqref{PR} neglects the explicit retroaction of the electron on the coherent-state amplitudes $\alpha$. One might expect this retroaction to be essential for modeling photon dynamics, yet the results reveal that it has only a minor impact on observables. Instead, the dominant mechanism driving the time evolution of the photon field is encoded in the entanglement structure of Eq.~\eqref{psi}. When the photon field is expressed in the coherent-state basis, its evolution is driven primarily by the electron-photon entanglement rather than the small electron-induced modification of the coherent parameter $\alpha$. This key insight explains why the approximate $R$-representation remains accurate while the $Q$-representation breaks down: the latter discards the off-diagonal correlations that encode this entanglement-mediated dynamic.


In summary, our full quantum TDSE simulations establish rigorous benchmarks for strong-field ionization driven by BSV light.
Our study shows that the $Q$-representation accurately describes total photoelectron spectra but fails to capture electron-photon entanglement signatures in joint energy spectra and photon-number-resolved observables. 
The breakdown originates from the absence of interference effects in the $Q$-representation, which leads to its inability to describe parity-dependent ionization pathways, as evidenced by pronounced even-odd photon-number modulations. Joint electron-photon spectra can be accessed using recent developments in coincident detection techniques, which already enable correlated measurements of photon statistics and emitted electrons in strong-field settings~\cite{needle}. 
This work bridges strong-field physics with quantum information science by establishing entanglement as a fundamental observable in strong-field quantum optics.


\textit{Acknowledgments}\textemdash
This work was supported by National Natural Science Foundation of China (NSFC) (Grant Nos. 12450405, 12274294, 12574378, 12574377, 11925405). The computations in this paper were run on the Siyuan-1 cluster supported by the Center for High Performance Computing at Shanghai Jiao Tong University. P.-L. H. is supported by the Yangyang Development Fund and the Pujiang Program of the Shanghai Baiyulan Talent Plan (Grant No.~24PJA046). Y.-J.M. is supported by T.D. Lee scholarship.

\textit{Data availability}\textemdash The data that support the findings of this work are openly available~\cite{data}.

\bibliography{ref}

\end{document}